\documentclass[twocolumn,showpacs,preprintnumbers,amsmath,amssymb,prb]{revtex4}

\usepackage{epsfig}

\begin{document}

\title{Hall effect in cobalt-doped TiO$_{2-\delta}$}

\author{J. S. Higgins}
 \email{sublime@glue.umd.edu}
\author{S. R. Shinde}
\author{S. B. Ogale}
\author{T. Venkatesan}
\author{R. L. Greene}
\affiliation{Center for Superconductivity Research, Department of
Physics, University of Maryland, College Park, Maryland
20742-4111}

\date{\today}

\begin{abstract}
We report Hall effect measurements on thin films of cobalt-doped
TiO$_{2-\delta}$. Films with low carrier concentrations (10$^{18}$
- 10$^{19}$) yield a linear behavior in the Hall data while those
having higher carrier concentrations (10$^{21}$ - 10$^{22}$)
display anomalous behavior near zero field.  In the entire range
of carrier concentration, n-type conduction is observed. The
appearance of the anomalous behavior is accompanied by a possible
structural change from rutile TiO$_{2}$ to Ti$_{n}$O$_{2n-1}$
Magn\'{e}li phase(s).
\end{abstract}

\pacs{75.50.Pp, 73.50.Jt, 73.50.-h}
\maketitle

In the field of spintronics, one of the major foci is the attempt
to inject spin-polarized current into existing semiconductor
technology, ultimately at room temperature (RT).  A possible
method is the use of magnetic semiconductors, unfortunately the
Curie temperatures (T$_c$) of these materials are significantly
lower than RT, resulting in little practical relevance.  Another
possibility is to take existing semiconductor materials and dope
them with magnetic impurities, called diluted magnetic
semiconductors (DMS).  The idea is to retain the parent compound's
semiconducting properties while adding ferromagnetism to the
system. Ga$_{1-x}$Mn$_x$As, the most extensively studied DMS,
exhibits T$_c$s as high as 160K,\cite{Chiba} which, while higher
than most, is still too low for practical purposes.

Recently, oxide DMS systems have shown ferromagnetism above RT.
One promising oxide is Ti$_{1-x}$M$_x$O$_2$ (M = magnetic
dopant).\cite{Matsumoto}  However, evidence shows that in the
anatase Co:TiO$_{2-\delta}$ system, clustering of cobalt atoms
occurs above a certain doping level (~2-3\%), and it is believed
that the observed high temperature ferromagnetism in such samples
is manifested in these clusters.\cite{Chambers,Shinde,Kim}  Under
specific growth and annealing conditions, samples without any
obvious clusters have also been shown to exhibit ferromagnetism
with a T$_{c}$ close to 700K.  However, whether the ferromagnetism
in this system is carrier-induced or extrinsic still remains an
unresolved issue. In this context, studies of the Hall effect,
Optical Magnetic Circular Dichroism (O-MCD), and electric field
effect measurements have been suggested to be the clarifying
experimental windows.  In this work, we report our observations on
the Hall effect in the Co:TiO$_{2-\delta}$ system.  While our work
was in progress, two groups reported on electronic transport
properties in oxide DMS systems.  Toyosaki \emph{et al.}
\cite{Toyosaki} reported an anomalous Hall effect in rutile
Co:TiO$_{2-\delta}$, and Wang \emph{et al.} \cite{Wang} found
similar effects in rutile Fe:TiO$_{2-\delta}$. These results are
suggested to imply that the observed ferromagnetism influences the
electronic transport in this material.

We grew thin films of anatase and rutile
Ti$_{1-x}$Co$_x$O$_{2-\delta}$ (x=0, 0.02) via pulsed laser
deposition.  The low cobalt concentration was chosen such that
cobalt clusters would be less likely to occur. We used
stoichiometric ceramic targets and deposited films through a Hall
bar shadow mask onto LaAlO$_3$ substrates (for anatase films) and
R-Al$_2$O$_3$ (1$\bar{1}$02) substrates (for rutile films). The
substrate heater temperature was 700 $^{o}$C and the laser energy
density was 1.8 $\frac{J}{cm^{2}}$ at 3 Hz. Magnetization
measurements were made using a Quantum Design SQUID magnetometer
and transport measurements were made using a Quantum Design
Physical Property Measurement System (PPMS).

Initially, we studied anatase Co:TiO$_{2-\delta}$ films.  In order
to obtain the anatase structure, we grew films on LaAlO$_{3}$ in
an oxygen environment of 10$^{-4}$ to 10$^{-8}$ Torr.  At higher
pressures (P$_{O2}$$\geqslant$10$^{-6}$Torr), the films grew in
(001) anatase form and showed RT ferromagnetic
behavior.\cite{Shinde}  However, in Hall measurements, we did not
observe an anomalous Hall effect (AHE). At lower pressures, the
anatase structure was compromised and gave x-ray diffraction (XRD)
scans different from the (001) anatase films.  From the peak
positions, it appeared to us that the film was rutile TiO$_{2}$.
Hall measurements on this film exhibited a small, non-linear
behavior near zero field (not shown). These results prompted
further investigation into highly oxygen deficient rutile films.

\begin{figure}
\centerline{\epsfig{file=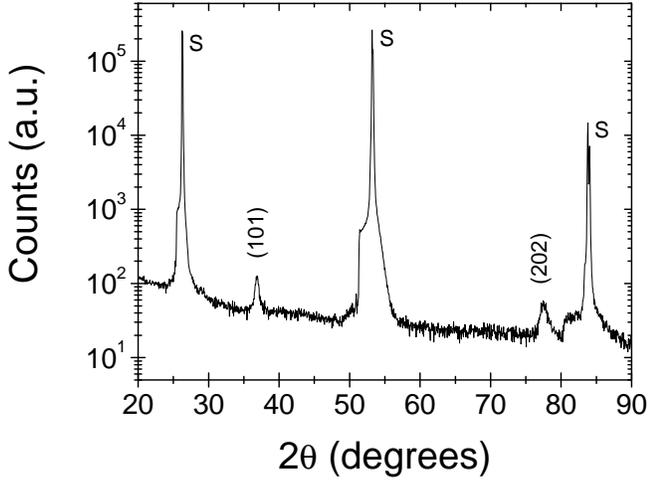,clip=,silent=,width=3.375in}}
\caption{XRD scan of sample 1.  The scan for sample 2 is nearly
identical.  The peaks labeled 'S' are substrate peaks.}
\label{Hall1}
\end{figure}

We used two approaches to increase oxygen vacancies in rutile
Co:TiO$_{2-\delta}$, as the conduction electrons originate from
these vacancies. Sample 1 was grown in vacuum with a base pressure
of 2x10$^{-8}$ Torr. Sample 2 was deposited using a 5\%
Hydrogen-Argon mixture at 1 mTorr of pressure. X-ray diffraction
(XRD), in Fig.~\ref{Hall1}, shows that sample 1 grew in the rutile
(101) structure.\cite{Wang,XRD}  Sample 2 showed similar XRD
patterns.

\begin{figure}
\centerline{\epsfig{file=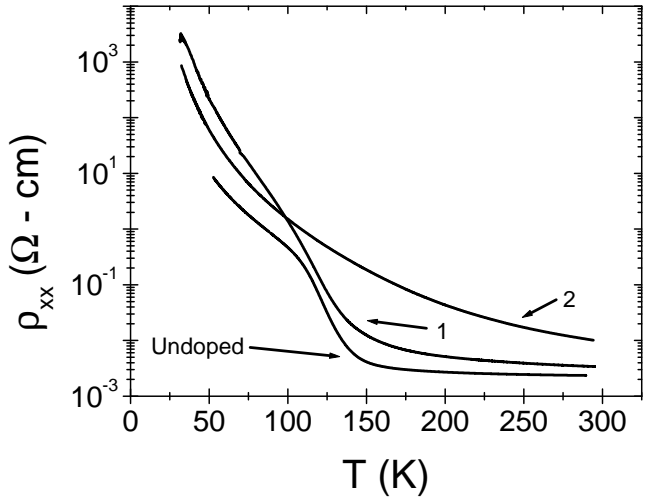,clip=,silent=,width=3.375in}}
\caption{Resistivity curves for sample 1, sample 2, and an undoped
film.} \label{Hall3}
\end{figure}

Both films display a relatively high conductivity ($\rho_{300K}$ =
2.53m$\Omega$-cm and 13.4m$\Omega$-cm for sample 1 and sample 2,
respectively), shown in Fig.~\ref{Hall3}.  As the temperature
decreases, the resistivity of sample 2 increases in an activated
manner whereas sample 1 shows an elbow near 140K. Similar behavior
was observed by Toyosaki \emph{et al.} \cite{Toyosaki} and Wang
\emph{et al.} \cite{Wang} for their films in which an AHE was
observed. The resistivity of sample 1 is not an expected result
due to the elbow, whereas sample 2 displays a temperature
dependence similar to bulk TiO$_{2-\delta}$.  The temperature
behavior of sample 1, however, matches more closely with the
Magn\'{e}li phase of this material (Ti$_{n}$O$_{2n-1}$).
\cite{Ueda}  This different phase of Ti-O orders in the rutile
structure of TiO$_{2-\delta}$, so XRD scans may not be able to
differentiate between the Magn\'{e}li phases and the rutile
TiO$_{2}$ phase. We also grew an undoped film in the same manner
as sample 1. This resistivity of this sample has a temperature
dependence similar to sample 1. Therefore, the temperature
behavior of the resistivity of our TiO$_{2-\delta}$ films is
influenced by the oxygen deficiency rather than the magnetic
dopant (cobalt).

The Hall effect arises from the Lorentz force deflecting charges
moving in a perpendicularly oriented magnetic field.  This
establishes an electric field transverse to the current. Typically
this effect is linear in field.  However, in magnetic materials,
the magnetic moment associated with the atoms gives rise to an
additive term in the Hall equation, \cite{Hurd,Chien}
\begin{equation}\label{1}
\rho_{xy}=\frac{E_{y}}{J_{x}}=R_{0}B+R_{A}\mu_{0}M_{S},
\end{equation}
where $\rho_{xy}$ is the Hall resistivity, E$_y$ is the electric
field perpendicular to the current and magnetic field, J$_x$ is
the current density, R$_0$ is the ordinary Hall coefficient, R$_A$
is the anomalous Hall coefficient, $\mu_0$ is the permeability of
free space, and M$_S$ is the field-dependent spontaneous
magnetization of the material. This anomalous Hall term is
conventionally attributed to asymmetric scattering processes
involving a spin-orbit interaction between the conduction
electrons and the magnetic moments in the material. At low
magnetic fields, the behavior of $\rho_{xy}$ is dominated by the
field dependence of M$_S$. Once the material's magnetization is
saturated, the $\rho_{xy}$ field dependence is linear and due to
the ordinary Hall effect. In many materials, R$_A$ shows a strong
temperature dependence, which usually correlates with the
electrical resistivity.

\begin{figure}
\centerline{\epsfig{file=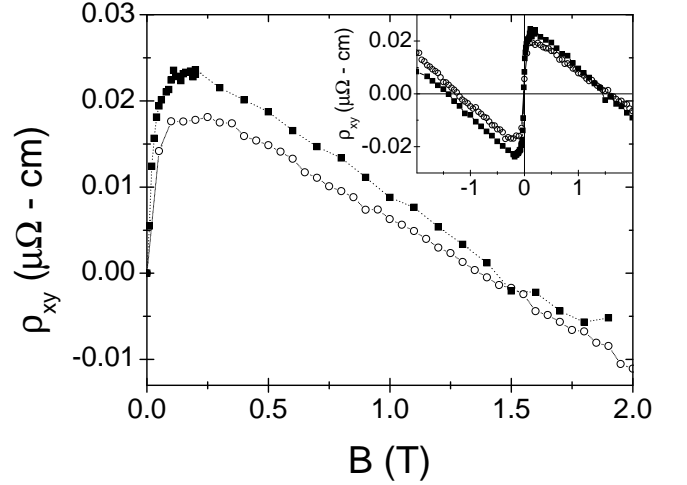,clip=,silent=,width=3.375in}}
\caption{Hall resistivity for sample 1.  Closed symbols are taken
at 300K, and open symbols are taken at 200K.  The respective
resistivities are 2.54 m$\Omega$-cm and 3.22 m$\Omega$-cm.}
\label{Hall4}
\end{figure}

\begin{figure}
\centerline{\epsfig{file=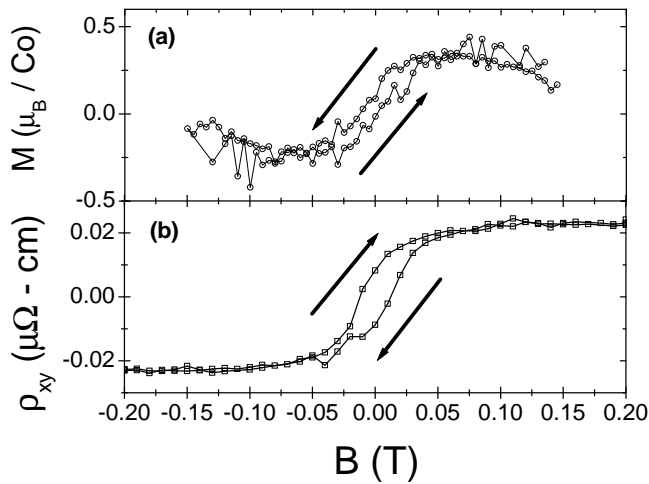,clip=,silent=,width=3.375in}}
\caption{(a) Magnetic hysteresis loop at 300K for sample 1. (b)
Expanded view of the Hall resistivity for sample 1 at 300K.}
\label{Hall4ab-2}
\end{figure}

\begin{figure}
\centerline{\epsfig{file=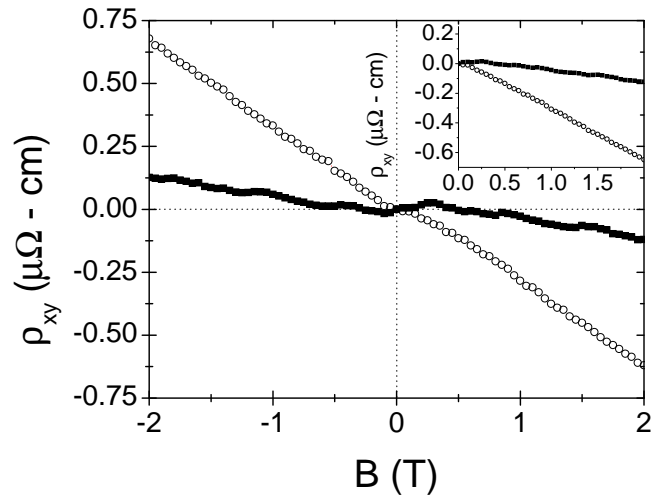,clip=,silent=,width=3.375in}}
\caption{Hall resistivity for sample 2 at 300K (closed symbols)
and 200K (open symbols).  The resistivities are 13.4 m$\Omega$-cm
and 57.2 m$\Omega$-cm respectively.} \label{Hall6}
\end{figure}

\begin{figure}
\centerline{\epsfig{file=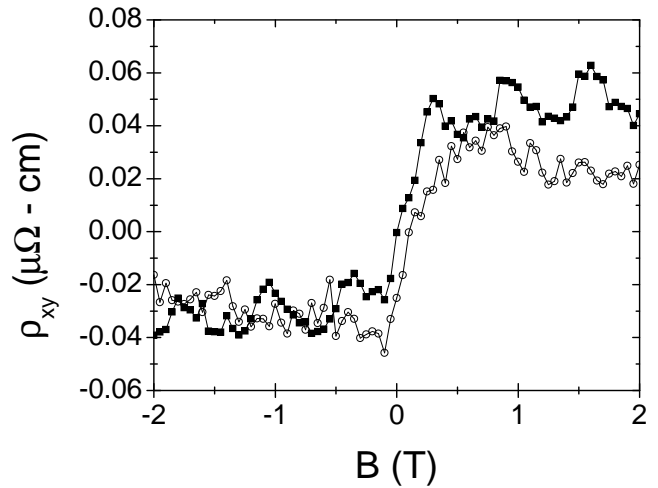,clip=,silent=,width=3.375in}}
\caption{Hall resistivity for sample 2 in which the normal Hall
contribution is subtracted from the data.  Closed symbols are
taken at 300K and open symbols are taken at 200K.} \label{Hall7}
\end{figure}

The field dependence of $\rho_{xy}$ for sample 1 is shown in Fig.
~\ref{Hall4}, measured at 300K and 200K.  The data were obtained
by a simple subtraction in order to eliminate any magnetic field
effects which are an even function of field, i.e.
magnetoresistance (MR)
($\rho_{xy}$=$\frac{1}{2}$[$\rho_{xy}$(H$^{+}$)-$\rho_{xy}$(H$^{-}$)]).
The inset shows the data before MR subtraction.  These data show a
sharp increase in $\rho_{xy}$ at low fields and a linear behavior
at higher fields, as expected for ferromagnetic materials. The
magnetic hysteresis loop for sample 1, measured with the field
perpendicular to the film plane, is shown in
Fig.~\ref{Hall4ab-2}(a).  For comparison, the Hall data is
expanded and replotted in Fig.~\ref{Hall4ab-2}(b).  The field at
which the magnetization saturates ($\sim$0.1 T) coincides well
with the low field behavior of the Hall data.  Therefore, the
rapid increase in $\rho_{xy}$ at low field can be interpreted as
an AHE. It is important to note that the negative slope of the
high field Hall data indicates n-type carriers. This is in
contrast with earlier reports, \cite{Toyosaki,Wang} but is
expected for TiO$_{2-\delta}$. The negative slope at high fields
gives an effective carrier concentration of
3.3$\pm$0.2x10$^{22}$/cm$^{3}$ at 300K and
3.56$\pm$0.02x10$^{22}$/cm$^{3}$ at 200K.  The Hall data for
sample 2 is shown in Fig.~\ref{Hall6}. The inset shows the data
after MR subtraction. A small but noticeable effect can be seen
around zero field. However, if we subtract the ordinary Hall
component from the data (determined from high fields), a clear
effect can be seen near the origin (Fig.~\ref{Hall7}).  As in
sample 1, sample 2 displays n-type behavior. The effective carrier
concentration is 8.0$\pm$0.1x10$^{21}$/cm$^{3}$ at 300K and
1.837$\pm$0.005x10$^{21}$/cm$^{3}$ at 200K.

The rather large carrier concentration observed in these highly
reduced samples raises some questions. It is known that oxygen
vacancies contribute shallow donor states in TiO$_{2-\delta}$. A
pure rutile film of TiO$_{2-\delta}$, grown by the same method as
sample 1, gave a carrier concentration of
3.09$\pm$0.02x10$^{22}$/cm$^{3}$ at RT, consistent with the
cobalt-doped samples. This observed carrier density would then
suggest the presence of approximately one oxygen vacancy for every
unit cell ($\delta$$\sim$0.5). This large carrier density, along
with the resistivity behavior, suggests that Magn\'{e}li phases
are present in films made using our growth conditions.

Our Hall measurements give clear evidence for an AHE in the
heavily oxygen reduced samples. Is this effect intrinsic to the
material, or is it a result of cobalt nanoclusters?  First, the
low field data changes behavior at nearly the same point that the
magnetization of the sample saturates. The magnetic saturation in
our films occurs at a field that is significantly lower than that
for cobalt metal films (H$\sim1.5-2T$). Second, since the
resistivities of each sample remain nearly the same for the two
temperatures measured, we expect the AHE to remain relatively
constant (in magnitude) for each sample, as is suggested by our
measurements.  While it is tempting to argue that the encouraging
observation of the AHE in the cobalt doped TiO$_{2-\delta}$ system
\emph{clearly} testifies to its carrier-induced or intrinsic
ferromagnetic character, other material-related possibilities
cannot be completely ruled out at this stage.  Specifically, the
question of cobalt clustering still lingers in view of the absence
of a clear theoretical negation of the occurrence of the AHE for
such cases. The structural and chemical microstructures formed in
samples prepared under highly reduced conditions could be quite
complex, especially in view of the known occurrence of Magn\'{e}li
phases in the oxygen-reduced Ti-O system.  Indeed, our preliminary
Transmission Electron Microscopy (TEM) observations on highly
reduced samples show the presence of some $\sim$10nm clusters at
the interface.  Kim \emph{et al.} \cite{Kim} have also observed
cobalt nanoclusters in their anatase Co:TiO$_{2-\delta}$ films
when the samples are grown in a low pressure oxygen environment
(~10$^{-7}$ Torr).  We propose to perform detailed studies on our
samples to examine the relative proportion of dissolved and
clustered cobalt to determine how the AHE could be interpreted in
these terms.

In summation, we have investigated electronic transport
measurements in the Ti$_{0.98}$Co$_{0.02}$O$_{2-\delta}$ DMS
system.  All films displayed n-type behavior and an increase in
carrier concentration with an increase in oxygen vacancies, which
are expected behaviors in the parent compound TiO$_{2-\delta}$. We
have found that, among several films grown at different oxygen
pressures and on different substrates, only the rutile films
exhibited an AHE when grown at low enough oxygen pressures.  In
spite of the observation of an AHE, it may be premature to
conclude that ferromagnetism in Co:TiO$_{2-\delta}$ is intrinsic.

Acknowledgement:  The author would like to thank V. N. Kulkarni
and H. Zheng for fruitful discussions.  This work was supported by
NSF under MRSEC Grant No. DMR-00-80008, and DARPA Spin-S
($\sharp$N000140210962) programs.


\begin{references}
\bibitem{XRD} The x-ray intensities of the (101) and (202) peaks
in the highly reduced films were found to be quite low as compared
to anatase and rutile films grown at relatively higher oxygen
pressures.  This could be due to the intergrowths of the different
Magn\'{e}li phases (Ti$_{n}$O$_{2n-1}$, n=4,5,6) in such samples.
\end{references}
\end{document}